# The possible origin of the higher magnetic phase transition in RuSr$_2$Eu$_{1.5}$Ce$_{0.5}$Cu$_2$O$_{10}$ (Ru-1222)

I. Felner, E. Galstyan, and I. Nowik

*Racah Institute of Physics, The Hebrew University, Jerusalem, 91904, Israel.*

Two magnetic transitions are observed in the magneto-superconducting RuEu$_{2-x}$Ce$_x$Sr$_2$Cu$_2$O$_{10-\delta}$ (Ru-1222), at $T_M \sim 160$ K and $T_{M2} \sim 80$ K. Below $T_{M2}$ the Ru moments are weak-ferromagnetically ordered and wide ferromagnetic hysteresis loops are observed, they become narrow and disappear at $\sim 60$-70 K. Above $T_{M2}$, (i) small antiferromagnetic-like hysteresis loops reappear with a peak in the coercive fields around 120 K. (ii) A small peak at $\sim 120$ K is also observed in the dc and ac susceptibility curves. The two phenomena are absent in the non-SC x=1 samples. For x<1, the decrease of the Ce$^{4+}$ content, is compensated by non-homogeneous oxygen depletion, which may induce a reduction of Ru$^{5+}$ ions to Ru$^{4+}$. The higher ordering temperature, $T_M$, which does not change with x, may result from Ru$^{4+}$ rich clusters, in which the Ru$^{4+}$-Ru$^{4+}$ exchange interactions are stronger than the Ru$^{5+}$-Ru$^{5+}$ interactions. In the superconducting Ru$_{1-x}$Mo$_x$Sr$_2$Eu$_{1.5}$Ce$_{0.5}$Cu$_2$O$_{10}$ (x=0-0.4) system, $T_{M2}$ shifts to low temperature with x (14 K for x=0.4), whereas $T_M$ is not affected by the Mo content, indicating again that $T_M$ may not correspond to the main phase. Two scenarios are suggested to explain the magnetic phenomena at $T_{M2} < T < T_M$. (i) They are due to a small fraction of nano-size islands inside the crystal grains, in which the Ru$^{4+}$ concentration is high and are magnetically ordered below $T_M$. (ii) The presence of nanoparticles of a foreign minor extra Ru$^{4+}$ magnetic phase of Sr-Cu-Ru-O$_3$, which orders at $T_M$, in which Cu is distributed inhomogeneously in both the Ru and Sr sites. This second scenario is supported by Mössbauer spectroscopy of $^{57}$Fe doped in Sr-Cu-Ru-O$_3$ systems.

## 1. INTRODUCTION

Coexistence of weak-ferromagnetism (W-FM) and superconductivity (SC) was discovered a few years ago in RuSr$_2$R$_{2-x}$Ce$_x$Cu$_2$O$_{10}$ (R=Eu and Gd, Ru-1222) layered cuprate systems[1-3], and more recently[4] in RuSr$_2$GdCu$_2$O$_8$ (Ru-1212). In both systems the SC and the magnetic states are confined to the CuO$_2$ planes and to the Ru layers respectively. The Ru-1222 materials display a magnetic transition at $T_M$= 130-180 K and bulk SC below $T_C$ = 32-50 K ($T_M > T_c$) depending on oxygen concentration and sample preparation. X-ray-absorption near- edge spectroscopy (XANES) measurements taken at the K edge of Ru, reveals that the Ru ions are basically pentavalent (4.95(5)) irrespective of the Ce concentration.[5] The *superconducting* state in both Ru-1212 and Ru-1222 systems is well established and understood. On the other hand, despite extensive research including neutron diffraction studies[6] the magnetic structure of the Ru-1212 system is far from being understood.[7] Moreover, the published data up to now have not included any determination of the magnetic structure in Ru-1222.

Generally speaking, the dc magnetic features of the Ru-1222 system exhibit two magnetic transitions, at $T_{M2}$ and at $T_M$ ($T_{M2} < T_M$). At low applied magnetic fields (H<5 kOe), irreversibility in the zero-field-cooled (ZFC) and field-cooled (FC) curves is observed, which disappears at $T_{M2}$. The dc ZFC curves (as well as the ac susceptibility plots) show a well-distinguished peak (denoted as the *first* peak) at temperatures ($T_P$), which are a bit lower than $T_{M2}$. For higher external fields (H>5 kOe), both ZFC and FC curves collapse to a single ferromagnetic-like behavior. A *second* small peak is observed around 120 K in both dc and ac susceptibility curves. $T_M$ was obtained from the temperature dependence of the saturation moment ($M_{sat}$), and from Mossbauer studies (MS) on $^{57}$Fe doped materials.[1] Wide FM hysteresis loops are opened at low temperatures (the coercive field- $H_C$ ~450-500 Oe at 5 K) which become narrower as the temperature increases and practically disappears around 60-70 K. Above $T_{M2}$, small field-induced canted AFM-like hysteresis loops are observed and the $H_C$(T) curves show a maximum around 120 K ($H_C$ ~150 Oe) and become zero at $T_M$.[8]

One of the disputed question in Ru-1222 materials, is the *origin* of the higher magnetic transition at $T_M$~160 K, whether it is intrinsic or not. The accumulated results were interpreted in the past by three scenarios. Scenario (A) also supported by ESR studies[9], assumes that the materials are *uniform as a whole* and that *all* the Ru ions behave in the same manner[4]: At $T_M$, the Ru sublattice becomes AFM ordered as mentioned above, and at $T_{M2}$ ($<T_M$), the whole material becomes W-FM ordered due to reorientation of the Ru moments, induced by the tilting of the $RuO_6$ octahedra from the crystallographic c axis.[10-11] At $T_C$~30 K the system as a whole becomes superconducting. Scenario (B) suggests phase separation.[12] Nano-size islands inside the crystal grains of the Ru-1222 materials, become FM at $T_M$, whereas the major part orders AFM at $T_{M2}$ and then becomes SC at $T_C$. In this scenario, the reopening of the hysteresis loops above $T_{M2}$, and the peak observed in $H_C$ cannot be reconciled, because the hysteresis loops opened at $T_M$ would remain all the way down to low temperatures and $H_C$ would increase continuously, or at least remain constant. Scenario (C), which is a combination of the two (A) and (B) models, also assumes phase separation but in a different manner. Our detailed MS analysis[13] on $^{57}$Fe and $^{119}$Sn doped materials, as well as muon spin rotation studies[14] indicate with little doubt the occurrence of two magnetic phases (with a minor phase of 15(5)%), which order magnetically at two different temperatures. Due to inhomogeneity of the oxygen content across the samples, a *minority* fraction (~10%) starts to order magnetically at $T_M$, displaying hysteresis loops. When the temperature is lowered, the AFM alignment of the spins strengthens. $H_C$ goes through a maximum, and then decreases and disappears. The *majority* fraction becomes W-FM ordered at $T_{M2}$ and displays the typical FM-like hysteresis loops.

During the course of studying the Ru-1222 system, we noticed that Ru can be replaced completely by Mo and that the $MoSr_2R_{1.5}Ce_{0.5}Cu_2O_{10}$ (Mo-1222) system is iso-structural with the Ru-1222 one. The Mo-1222 system can be obtained with most of the R elements (Pr-Yb and Y)[15]. In contrast to the Ru-1222 described above, $MoSr_2Eu_{1.5}Ce_{0.5}Cu_2O_{10}$, is SC only, with $T_C = 23$ K.

In order to elucidate the origin of the second peak and the reopening of the hysteresis loops at $T_{M2}<T< T_M$, we first report here data of the mixed $Ru_{1-x}Mo_xSr_2Eu_{1.5}Ce_{0.5}Cu_2O_{10}$ (x=0.1-1). Keeping in mind that such a substitution dilutes the magnetic $RuO_2$ layers contribution, we show here that Mo substitution (up to x=0.4), shifts only $T_{M2}$ ($T_P$) to lower temperatures, but does not alter much $T_M$ and the second peak position. One may argue that the higher magnetic transition is due to the FM $SrRuO_3$ impurity phase, which orders at 165 K. However, the comparison between the MS of $^{57}$Fe doped in Ru-1222 and in *pure* $SrRuO_3$ materials, excludes this assumption.[16] We show here that in the mixed $Sr-Cu-Ru-O_3$ systems, (Cu is substituted for either Ru or Sr), a pronounced peak in the dc $H_C(T)$ plots around 120 K is obtained, similar to that observed in Ru-1222. Moreover, the MS spectra of $^{57}$Fe doped in the $Sr_{1-x}Cu_xRuO_3$ system, (which are completely different from that of pure $SrRuO_3$), are reminiscent of the MS of Fe doped in Ru-1222. Thus, we argue that the higher magnetic transition is due to the presence of scattered islands (nano-particles) of the mixed $Sr-Cu-Ru-O_3$ phase as an impurity phase. Alternatively, we suggest that in $RuSr_2R_{2-x}Ce_xCu_2O_{10-\delta}$, for x<1, due to oxygen deficiency ($\delta$) and/or inhomogeneity in the oxygen content, a *minor fraction* of the $Ru^{5+}$ ions is reduced to $Ru^{4+}$ islands. These $Ru^{4+}$ rich nano-size islands order at $T_M$, while the *major* $Ru^{+5}$ bulk material orders at $T_{M2}$, thus the two magnetic transitions in Ru-1222, are related to two intrinsic magnetic regions. Both options agree well with scenario (C) described above.

## 2. EXPERIMENTAL DETAILS

Ceramic samples with nominal composition $Ru_{1-x}Mo_xSr_2Eu_{1.5}Ce_{0.5}Cu_2O_{10}$ (x=0.1-1) or Sr-Cu-Ru-$O_3$ were prepared by a solid-state reaction technique. Prescribed amounts of $Eu_2O_3$, $CeO_2$, $SrCO_3$, CuO, Ru, and Mo were mixed and pressed into pellets and preheated at 950° C for 1 day. The as prepared (asp) products were cooled, reground and sintered at 1050° C for 2 days (1200° C for Sr-Cu-Ru-$O_3$), reground and heated at 1050° C for 2 days under an oxygen atmosphere and then furnace cooled. The materials characterized as reheated were heated two more days under the same conditions. Each series of the materials was prepared *at the same time under the same conditions*. X-ray diffraction (XRD) measurements confirmed the purity of the compounds. Dc magnetic measurements were performed in a commercial (Quantum Design) super-conducting quantum interference device (SQUID) magnetometer. Ac susceptibility was measured at $H_{dc}=0$ (zero applied magnetic field) by a home made probe inserted in the SQUID, with an excitation frequency of 733 Hz and amplitude of 120 mOe. The microstructure and the phase integrity of the materials were investigated by QUANTA (Fri Company) scanning electron microscopy (SEM) and by a Genesis energy dispersive x-ray analysis (EDAX) device attached to the SEM. Mössbauer studies of ceramic samples containing ~1% $^{57}$Fe were performed at various temperatures using a conventional constant acceleration drive and a 50 mCi $^{57}$Co:Rh source. The experimental spectra were analyzed by a least square fitting procedure in terms of several sub-spectra, corresponding to various models based on distributions of hyperfine interaction parameters. The $^{57}$Fe isomer shifts are relative to α-Fe, measured at room temperature.

## 3. EXPERIMENTAL RESULTS

**(a) The mixed $Ru_{1-x}Mo_xSr_2Eu_{1.5}Ce_{0.5}Cu_2O_{10}$ system.**

All materials reported in this section were prepared *at the same time under the same conditions*. The Mo valence is not known as yet and our XAS experiments to determine it were unsuccessful, because the Mo valence is not stable under experimental conditions (e.g. the high vacuum in the chamber) and changes during the experiment. All the mixed $Ru_{1-x}Mo_xSr_2Eu_{1.5}Ce_{0.5}Cu_2O_{10}$ samples have a tetragonal structure (space group I4/mmm). Due to the similarity in the lattice parameters of both end compounds, $RuEu_{1.5}Ce_{0.5}Sr_2Cu_2O_{10}$ (a=3.841Å and c=28.38(2) Å and $MoEu_{1.5}Ce_{0.5}Sr_2Cu_2O_{10}$ a=3.838(1)Å and c=28.41(2)Å), the lattice parameters (within the instrumental accuracy), of all mixed materials are: a=3.840(1) and c=28.40(2) Å independent of the Mo content. The least squares fits of the XRD patterns of the mixed compounds, left a few minor reflections, most of them belonging to the Pauli-paramagnetic $SrMoO_4$ phase.[17] Detailed synchrotron X-rays diffraction study indicates, that the rotation of $RuO_6$ octahedra ~ 14° around the c-axis discovered in Ru-1222 (x=0)[9], essentially exists also in the x=0.6 and x=1 compounds.[18] The morphology detected by the SEM measured on several granular materials, shows a smooth and uniform surface, with typical grain size of 2-3 μm. EDAX analysis confirms the initial stoichiometric composition of Ru:Eu:Ce:Sr:Cu, whereas a deficiency in the Mo content is obtained due to its volatility. We also detected a few separate spherical grains of $SrMoO_4$. All the mixed materials are SC with $T_C$ values in the range of 27-19 K. Since the main aim of the present paper is the magnetic behavior and more specifically the origin of the second peak, the SC properties of these materials will not be discussed.

Macroscopic coexistence of SC and magnetism appears for samples with high Ru content (x<0.5), while the materials with x>0.5 are SC only. $Ru_{0.5}Mo_{0.5}Eu_{1.5}Ce_{0.5}Sr_2Cu_2O_{10}$ (x=0.5) is on the verge of the magnetic state. The asp sample is magnetically ordered and not SC, whereas the reheated one is SC only ($T_C$=25 K). The temperature dependence of the normalized ac susceptibility curves (at $H_{dc}$=0) for the reheated mixed $Ru_{1-x}Mo_x Eu_{1.5}Ce_{0.5}Sr_2Cu_2O_{10}$ (x<0.5) system is presented in Figs. 1 and 2. The main magnetic peak position is shifted to lower temperatures with increasing x: $T_P$ = 79 K for x=0 and 14 K for x=0.4 (see Fig. 2). On the other hand, $T_M$ for all samples defined by $M_{sat}$ (see above), does not alter much (~ 150-160 K). Note, the second peak around 116 K for x=0 and x=0.4 (Fig.1 main frame and inset) remains practically unchanged. Similar behavior was observed when Al was substitution for Ru (up to 10%).[13] The broad SC transitions are typical of inhomogeneous Cu based high $T_C$ materials, where inhomogeniety in the oxygen concentration causes a distribution in the $T_C$ values. Similar broadenings are observed by our dc resistivity measurements (not shown).

The M(H) curves are strongly dependent on the field, until a common slope is reached. At low applied fields, the M(H) curve exhibits a typical ferromagnetic-like hysteresis loop (Fig. 3) similar to that reported in Ref. 1 and 2. Fig. 3 (inset) shows the apparent tendency toward saturation at 5 K, without reaching full saturation even at 50 kOe. The M(H) curves (above H=10 kOe) can be described as: M(H)= $M_{sat}$+ $\chi$H, where $M_{sat}$ corresponds to the W-FM contribution of the Ru sublattice, and $\chi$H is the linear paramagnetic contribution of Eu and Cu. Fig. 2 (inset) shows the linear decrease of $M_{sat}$ with x (0.58(2)$\mu_B$ for x=0 at 5 K), indicating the dilution of the magnetic $Ru^{5+}$ ions by Mo ions.

Above $T_M$, the $\chi$(T) curves (measured at 10 kOe) for x=0-0.4, have the typical paramagnetic shape and adhere closely to the Curie-Weiss (CW) law: $\chi(T) =\chi_0 +C/(T-\theta)$, where $\chi_0$ is the temperature independent part of $\chi$, C is the Curie constant, and $\theta$ is the CW temperature. Since the $Eu^{3+}$ and $Cu^{2+}$ and Mo ions, all contribute to C and due to uncertainty in the Mo ions valence, the net Ru contributions to $\chi$(T) were not calculated. On the other hand, the $\theta$ values decrease sharply with x, $\theta$= 114 K for x=0 and 56 K for x=0.4, in full agreement with the shift of $T_P$ (Fig. 1). For x> 0.5, the $\chi$(T) curves do not follow the CW law.

## (b) $RuEu_{2-x}Ce_xSr_2Cu_2O_{10-d}$ (Ru-1222) system

The SC and magnetic phase diagram of the $RuEu_{2-x}Ce_xSr_2Cu_2O_{10}$ system was presented and discussed in our previous publication.[2] Generally speaking,

RuEuCeSr$_2$Cu$_2$O$_{10}$ (x=1) is magnetically ordered only, and SC appears for samples with x=0.8-0.4. Due to the similarity of the ionic radii of Eu$^{3+}$ (0.94 Å) and Ce$^{4+}$(0.87Å) the lattice parameters of all materials studied are basically unchanged.[2] We shall focus here on the new accumulated magnetic measurements, which concern the main topic of this paper.

The ZFC plots (measured at 4 Oe) of the asp RuEu$_{2-x}$Ce$_x$Sr$_2$Cu$_2$O$_{10-\delta}$ (x=1-0.4) materials are shown in Fig. 4. The ac curves show a similar behavior. The major peak for x=1 at $T_P$ =108 K is shifted to 75 K for x=0.4. For RuEuCeSr$_2$Cu$_2$O$_{10}$ (x=1), neither $T_P$ nor $T_M$ =160 K are altered by further annealing or quenching of the sample. On the other hand, reheating the x=0.8-0.4 asp materials, shifts $T_P$ to higher temperatures (85 K for x=0.4) whereas quenching the materials from 1050° C to ambient temperature shifts the peak to lower temperatures (70 K for x=0.4), a trend which is consistent with our previous results.[19] The second peak (around 120 K) appears clearly in the SC samples (x=0.8-0.4), in both ZFC and ac curves (Fig. 4 inset), whereas *no second peak* is observed whatsoever for the x=1 materials. Fig. 5 shows the $M_{sat}$(T) for the various samples. The information deduced from Fig. 5 is: (i) All $M_{sat}$ curves have a typical ferromagnetic like behavior and merge at $T_M$ =160 K. (ii) The $M_{sat}$ values gradually decrease with x. At 5 K, $M_{sat}$ for x=1 and x=0.4 are 0.88$\mu_B$ and 0.39$\mu_B$ respectively. For x=1, $M_{sat}$ remains unchanged for both reheated and quenched samples.

Wide ferromagnetic hysteresis loops exist at low temperatures, from which the coercive field ($H_C$) and the remnant moments can be deduced. The $H_C$ values depend also on x and **vary** (at 5 K), from 460 Oe for x=0.4 (Fig. 6) to ~320 Oe for x=0.8 and 1. As the temperature is raised the hysteresis loops become narrower and close themselves at ~60 K ($H_C$=0), thus essentially no discernible hysteresis exists above 60 K (Fig. 6). At T>80 K reappearance of $H_C$ is observed for x=0.4-0.8 and the temperature dependence of $H_C$ exhibits a bell shape behavior with a peak around 110-120 K (close to the peaks observed in Figs. 1, and 4). $H_C$ disappears at $T_M$, similar to the trend of $M_{sat}$ shown in Fig. 5. For x=1, the $H_C$ values for both the asp and quenched samples are the same (Fig. 6 inset), and no reappearance of $H_C$ at higher temperatures is observed. In contrast to the FM-like hysteresis loop obtained at T< 60 K (Fig. 3), the hysteresis loops above 90 K exhibit AFM like features.[8]

Fig.6 leads to the following conclusions: (i) The susceptibility second peak is connected with the bell shape of $H_C$. For x=1, neither a reappearance of $H_C$ nor a second peak are observed (ii) The general trend is that the maximum in $H_C$ (Fig. 6) and the second peak, both increase with the decrease of x. Indeed, the x=1 sample may serve as the parent stoichiometric insulator compound.[2] MS studies on $^{151}$Eu and XAS taken at $L_{III}$ and K edges of Ce and Ru respectively clearly indicate that the valence of $Eu^{3+}$, $Ce^{4+}$, $Ru^5$ are fixed and conclusive.[20] A straightforward valence counting (assuming $Sr^{2+}$, $Cu^2$ and $O^{2-}$) yields a fixed oxygen concentration of 10, which does not change either by further annealing nor by quenching the sample. As a result the magnetic parameters such as: $M_{sat}$ and $H_C$ are the same for all three (asp, reheated and quenched) materials.

The main interest now is in the *origin* of the two connected magnetic phenomena namely: the second peak and the peculiar behavior of $H_C$. The intriguing question arises as to why both observations do not appear in the x=1 material. In the absence of microscopic information on the magnetic structure, an interpretation of these anomalies is not straightforward. We suggest a model that can lead to the observed behavior. A central assumption is that the hole doping of the Cu-O planes, which results in metallic behavior and SC in $RuEu_{2-x}Ce_xSr_2Cu_2O_{10-\delta}$ for the x<1 materials, is obtained when trivalent $Eu^{3+}$ ions are replaced by $Ce^{4+}$. Reducing the $Ce^{4+}$ content is partially compensated by depletion of oxygen; thus the oxygen deficiency increases with $Eu^{3+}$. It is possible that this depletion is not homogeneous throughout the material and one cannot exclude tiny differences in the oxygen content in various clusters. There are clusters in which the deficiency is high, and in order to maintain neutrality, those $Ru^{5+}$ ions (say about 5-10%) that are surrounded by less oxygen as first neighbors are reduced to $Ru^{4+}$. This is consistent with the average Ru valence of 4.95(5) reported in Ref. 5. The $Ru^{4+}$- $Ru^{4+}$ exchange interactions (S=1) are stronger than the $Ru^{5+}$-$Ru^{5+}$ (S=1/2) and $Ru^{4+}$-$Ru^{5+}$ interactions, and cause these clusters to order magnetically at a higher temperature. According to this picture, the two magnetic transitions in Ru-1222, are due to different *intrinsic regions*, which differ in their oxygen distribution. The major W-FM fraction is one in which $Ru^{5+}$ orders at $T_{M2}$ (80-90 K) whereas the minor fraction of the material in which the Ru ions are basically tetravalent and order at $T_M$ (around 160 K). Here, two types of interactions are present. The relatively strong $Ru^{4+}$- $Ru^{4+}$ fraction orders at $T_M$, whereas the $Ru^{5+}$- $Ru^{4+}$ fraction starts to order at somewhat lower temperatures (higher than $T_{M2}$) and is probably ferrimagnetically coupled to the first one. This explains the AFM-like hysteresis loops as well as the peak in the magnetization (Fig. 4) observed at $T_{M2}<T< T_M$. It is reminiscent of the general trend in the ruthenates, in which materials with $Ru^4$ ions ($SrRuO_3$) have higher magnetic transitions than materials with $Ru^{5+}$ ions (e.g. $EuSr_2RuO_6$ $T_M$=30 K)[16].

This scenario explains well the results obtained in the mixed $Ru_{1-x}Mo_xSr_2Eu_{1.5}Ce_{0.5}Cu_2O_{10}$ system presented in Figs. 1-2. The ionic radius of $Ru^{5+}$ (0.71 Å) is smaller than that of $Ru^{4+}$ (0.76 Å). The smaller Mo ions (whether $Mo^{6+}$ (0.55Å) or $Mo^{5+}$ (0.75Å) substitute preferentially the $Ru^{5+}$ sites, dilute the $Ru^{5+}$-$Ru^{5+}$ interactions in the major W-FM fraction, and shift $T_{M2}$ dramatically to lower temperatures. Therefore, the presence of Mo does not alter the minor second peak position and $T_M$ of the minor fraction.

**(c) The Sr-Ru-Cu-$O_3$ system**

In view of the results presented so far, it appears tempting to argue that the higher magnetic transition in Ru-1222 and the anomalies shown in Figs. 5-6 are due to the itinerant ferromagnetic $SrRuO_3$ ($T_M$ =165 K), which may exist as an extra impurity phase. However, the comparison between the MS of Fe doped Ru-1222 and $SrRuO_3$ samples exclude the assumption raised above[1,16]. In Ru-1222, the dilute Fe probe, successfully follows the Ru magnetization and well-defined magnetic hyperfine fields are visible in the MS spectra, up to $T_M$.[1] On the other hand, in the MS spectra of $SrRuO_3$ at T>90 K, there is no sign whatsoever of magnetic order (in full agreement with Ref. 21), indicating a Fe–Ru weak coupling at elevated temperatures.

The FM behavior of $SrRuO_3$ was studied extensively and it is attributed to the highly correlated Ru 4d-electron band [22-27]. Since the Ru-1222 materials contain the elements Ru, Sr, and Cu, we decided to study the magnetic properties of the $SrRu_{1-y}Cu_yO_3$ and the $Sr_{1-x}Cu_xRuO_3$ systems in which Cu is substituted for Ru and Sr respectively. The materials were doped with 1% $^{57}$Fe to enable MS studies. The magnetic and MS studies of the two systems are quite different. In the first system, doping of $Cu^{2+}$ for Ru profoundly alters $T_M$ of $SrRuO_3$, whereas, in $Sr_{1-x}Cu_xRuO_3$ the doping does not change $T_M$ much. The MS spectra are also significantly different because of the difference in the Fe-Ru distance, which is a dominant factor.

FC magnetization curves for $Sr_{0.8}Cu_{0.2}RuO_3$ (open symbols) and for $SrRu_{1-y}Cu_yO_3$ (y=0.1, 0.2 and 0.3) are exhibited in Figs. 7. All FC curves have the typical FM shape and resemble the FM features of $SrRuO_3$ (not shown). Fig. 7 shows a sharp decrease of $T_M$ for $SrRu_{1-y}Cu_yO_3$ with the Cu content: 165 K (for y=0) to 160, 130 and 75 K for y= 0.1, 0.2 and 0.3 respectively. Those $T_M$'s are somewhat higher than the values published in Ref. 27. For higher Cu concentrations, a very inhomogeneous cationic distribution is observed and the samples were not investigated further.[26] In $Sr_{1-x}Cu_xRuO_3$, the decrease in the magnetic transition is more moderate and $T_M$ shifts to 150 and 147 K for x=0.1 and 0.2 respectively.

All isothermal M(H) curves below $T_M$, are strongly dependent on the field (up to 7-10 kOe), until a common slope is reached. At low applied fields, the M(H) curve exhibits a typical FM-like hysteresis loop. Fig. 8 shows the temperature dependence of $H_C$ for various samples investigated. For y=0.1, $H_C$ decreases rapidly with T up to 70 K, then increases with a peak at 100 and becomes zero at $T_M$. For y=0.2 a small peak is obtained at 110 K. For $Sr_{0.9}Cu_{0.1}RuO_3$ the peak in $H_C$ exists, but it is smeared and not so pronounced. Note that peaks in $H_C$ appear at the *same temperature range* as the peak in $H_C$ and the second magnetic peak in Ru-1222 presented in Figs.1, 4 and 6.

Mössbauer studies of dilute $^{57}$Fe probes have proved to be a powerful tool in the determination of the magnetic nature of the probe's site location. When the Ru ions become magnetically ordered, they produce an exchange field on the Fe ions residing in this site. The $^{57}$Fe nuclei experience magnetic hyperfine fields leading to sextets in the observed MS spectra. MS measurements performed at 90 K, on 1% $^{57}$Fe doped $Sr_{0.9}Cu_{0.1}RuO_3$ and $SrRu_{0.9}Cu_{0.1}O_3$ materials (Fig. 9) exhibit a significant difference between the two spectra. For the sake of comparison the 90 K MS of Ru-1222 is also shown. Due to the similarity in the ionic radii of $Cu^{2+}$ and $Fe^{3+}$, it is assumed that the Cu ions drag the probe Fe ions, which prefer to reside together with Cu at the same site. The MS spectrum of $SrRu_{0.9}Cu_{0.1}O_3$ ($T_M$=160 K) shows only one singlet, very similar to that of $SrRuO_3$.[16,21] Due to the large Ru-Ru (~5.55Å) distance, the weak Fe-Ru exchange interactions fail to probe the Ru magnetic order. On the other hand for $Sr_{0.9}Cu_{0.1}RuO_3$, two sub-spectra are observed at 90 K. A nonmagnetic singlet (~75%), and a magnetic sextet (~25%) which does follow the Ru magnetization and disappears at $T_M$. The magnetic sub-spectrum was fitted with a distribution of magnetic hyperfine fields with an average value of $H_{eff}$ =313(9) kOe. The average Sr-Ru distance (~2.78Å) is much shorter than the Ru-Ru one. This minor sextet results from those $Fe^{3+}$ ions, which reside in the mixed Sr-Cu site, and experience an exchange field from their *magnetic* $Ru^{4+}$ close neighbors. The major non-magnetic singlet corresponds to Fe ions that reside in the Ru site (like in pure $SrRuO_3$).

The main issue is the similarity between the MS spectra of $Sr_{0.9}Cu_{0.1}RuO_3$ and Ru-1222 spectra (Fig. 9). Thus, we assume that the second magnetic phase in Ru-1222 may arise from a small extra phase of magnetic $Sr-Cu-Ru-O_3$ phase, possible as small nano-domain particles, undetected by XRD. The phase composition and its fraction depend on the Ce and/or oxygen concentration as well as on the preparation procedures.

## 4. DISCUSSION and CONCLUSIONS

The results shown here are not compatible with both models (A) and (B) described in the introduction. In the absence of microscopic information, an interpretation of: (i) the small second peak in the magnetization curves, and (ii) the rise in $H_C$ at 120 K is not straightforward. We assume that the two phenomena are connected to each other and have the same origin. As stated above, in the stoichiometric parent $RuEuCeSr_2Cu_2O_{10}$ (x=1) compound none of those peaks are observed. They appear only for samples with less $Ce^{4+}$ (x<1) where the reduction of Ce content, is compensated for by depletion of oxygen, which is not homogeneous throughout the whole material. Those $Ru^{5+}$ ions that are surrounded by less oxygen as first nearest neighbors may reduce to $Ru^{4+}$, and we may assume, that the two phenomena measured above are related to this small fraction (which increases with Eu content) of the $Ru^{4+}$ ions. We suggest two scenarios that could lead to the observed behavior.

(i) It is an *intrinsic* bulk property. The $Ru^{4+}$- $Ru^{4+}$ exchange interactions are stronger than the $Ru^{5+}$-$Ru^{5+}$ ones and have a higher magnetic transition. A small fraction of nano-size islands species inside the crystal grains of Ru-1222 in which the $Ru^{4+}$ concentration is high, become magnetically ordered at $T_M$ (~ 160 K). At $T_{M2}$ (80-90 K), a W-FM is induced in the major part of the material, which originates from canting of the Ru moments, which arises from the Dzyaloshinsky-Moriya anti-symmetric super-exchange interaction.[1] This scenario is consistent with previous MS [13] and μSR [14] studies.

In order to calculate the minor fraction amount, we measured both $RuEu_{1.5}Ce_{0.5}Sr_2Cu_2O_{10-\delta}$ and the parent (x=1) compounds in the following procedure. (a) The samples were ZFC to 5 K and then measured at 6 Oe (or at 50 Oe) up to above $T_P$ of the materials, 99 K and 125 K respectively. (b) The samples were FC to 5 K and measured up to 185 K. (c) The samples were FC from 185 K to 5 K and measured again up to 185 K. The three curves measured at 6

Oe are exhibited in Fig. 10. One definitely observes the difference between the two FC branches. On the other hand, for the x=1 sample, no difference (within the experimental error) between the two FC processes is observed. The inset of Fig. 10 presents the careful subtraction of the two FC curves (measured at 50 Oe) for x=0.5 and 1, materials. For x=0.5, relatively high values at low temperatures and a pronounced peak at 105 K are observed, whereas, for x=1 the values are around zero for almost the entire temperature range. The contribution of the minor fraction for x=0.5 (at 5 K), to the FC magnetization is given by: $(FC_{185K}-FC_{99K})/FC_{99K}$ = 0.045. This value is consistent with XANES[5] measurements on $RuGd_{2-x}Ce_xSr_2Cu_2O_{10-\delta}$ (x=0.6-0.8) where an average Ru valence of 4.95(5) was extracted, which means that 95% Ru ions are $Ru^{5+}$ and the rest appear as $Ru^{4+}$. Another quantitative XANES study on asp and 100-atm. oxygen annealed materials revealed a Ru valence value of 4.74 and 4.81 respectively, confirming our present model that the average Ru valence is affected by the change in oxygen content.[28] Qualitatively speaking, the Ru-1222 system is somewhat similar to the Ru-1212 one in which NMR experiments suggest that the Ru ions may be in a mixed valence state with 40% $Ru^{4+}$ (S=1) and 60% $Ru^{5+}$.[29]

(ii) The alternative interpretation assumes the possibility that the higher magnetic transition is due to the presence of scattered islands of the mixed $Sr-Cu-Ru-O_3$ phase as an impurity phase[26], its composition is sample dependent. This phase also exhibits (a) a typical increase in $H_C$ around 120 K (Fig. 8), (its origin is now under investigation). A similar rise in $H_C$, was also detected in $Sr_{0.6}Ca_{0.4}RuO_3$.[25] (b) Due to the short Ru-Sr distance, those $^{57}Fe$ probe ions which reside in the Sr site, sense the magnetization of Ru up to $T_M$, unlike those which reside in the Ru site (Fig. 9). This phase (5-10%) may exist in the Ru-1222 samples (except for x=1) in which the oxygen depletion reduces the Ru valence to $Ru^{4+}$. The major part of the material becomes W-FM ordered at $T_{M2}$ and coexists with the SC state induced below $T_C$. Neutron diffraction measurements are required to precisely determine the properties of the magnetic order in the Ru-1222 system.

**Acknowledgments:** This research was supported by The Israel Science Foundation (grant No. 618/04), and by the Klachky Foundation for Superconductivity.

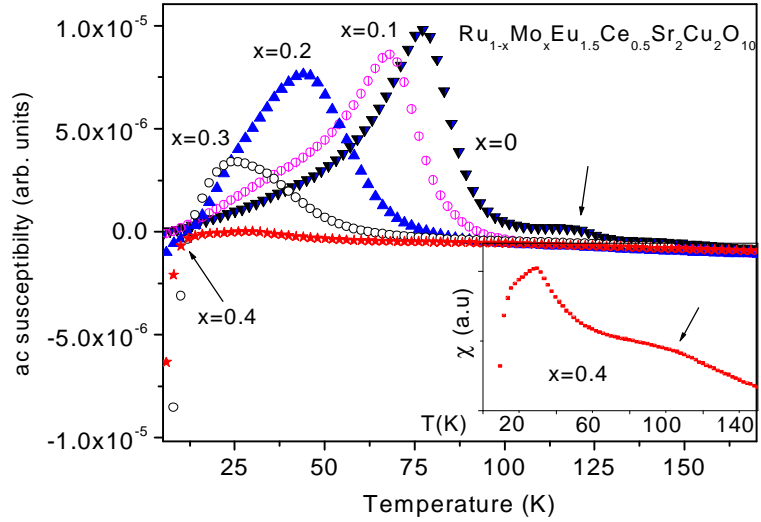

FIG. 1. Normalized ac susceptibility of $Ru_{1-x}Mo_xSr_2Eu_{1.5}Ce_{0.5}Cu_2O_{10}$. The inset shows the ac susceptibility of x=0.4 material in an extended scale.

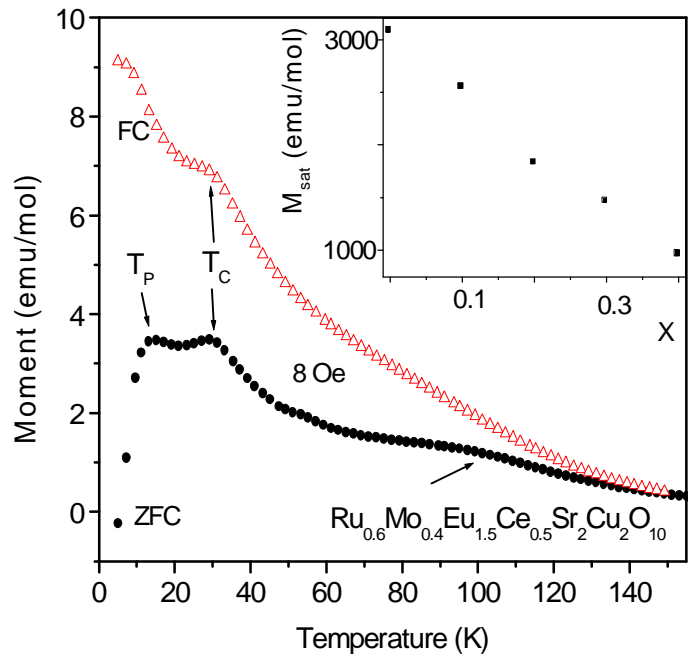

FIG. 2. ZFC and FC magnetization curves of $Ru_{0.6}Mo_{0.4}Sr_2Eu_{1.5}Ce_{0.5}Cu_2O_{10}$. Note the $T_P$ and $T_C$ temperatures and the second peak position. The inset shows the linear decrease of $M_{sat}$ (at 5 K) with x

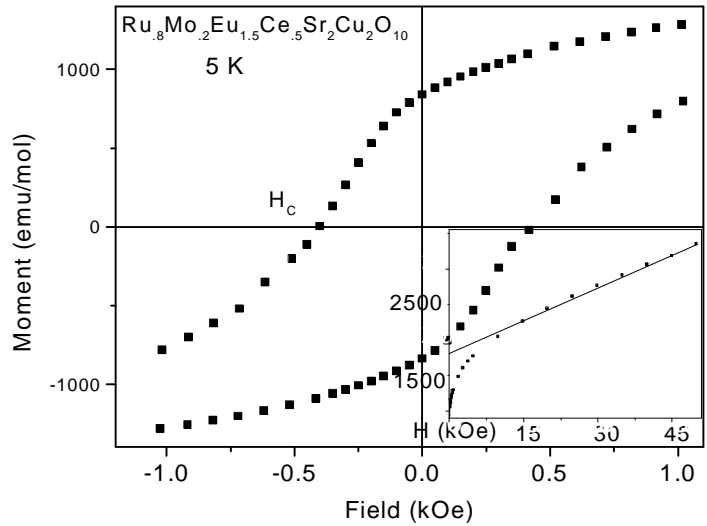

FIG. 3. The hysteresis loop at 5 K for $Ru_{0.8}Mo_{0.2}Sr_2Eu_{1.5}Ce_{0.5}Cu_2O_{10}$ and the typical M(H) curve up to 50 kOe from which $M_{sat}$ is deduced.

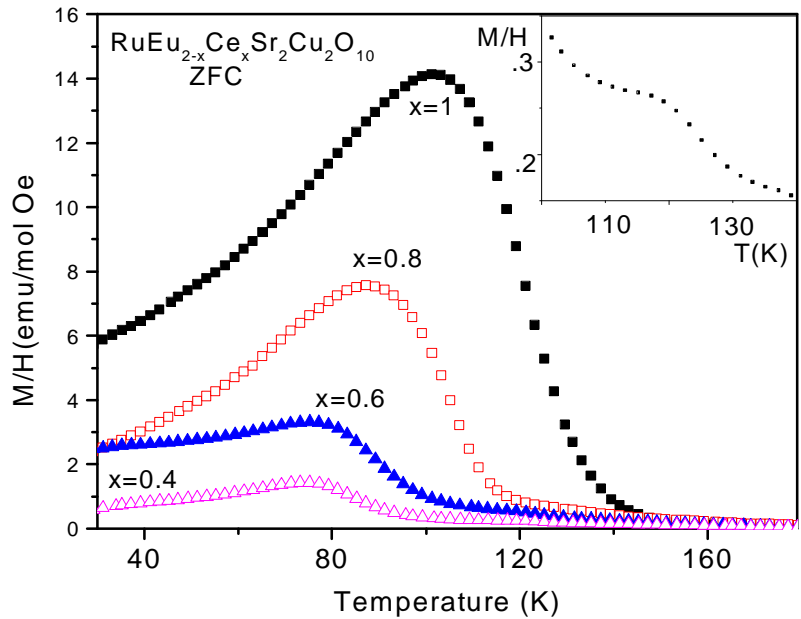

FIG. 4. ZFC susceptibility curves measured at 4 Oe for the asp $RuEu_{2-x}Ce_xSr_2Cu_2O_{10}$ materials. The inset shows the expanded high temperature range for x=0.4.

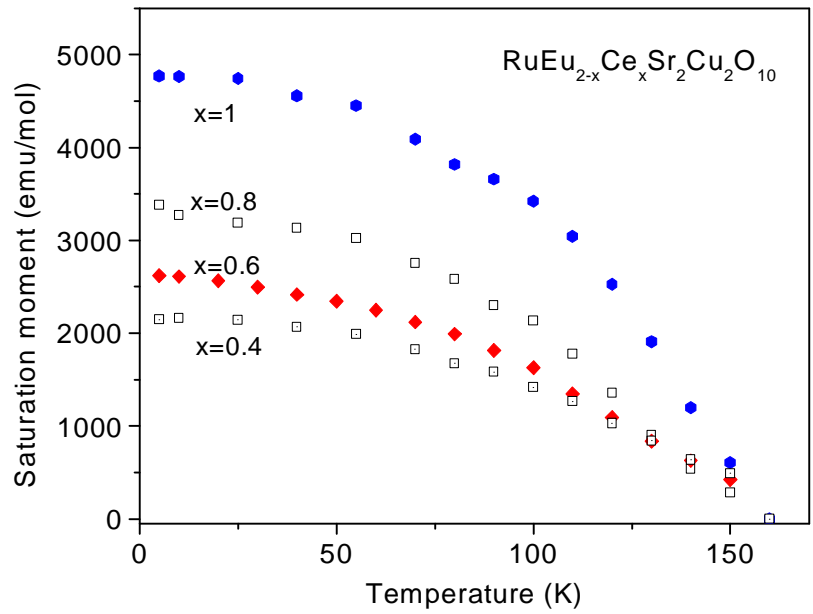

FIG.5. Temperature dependence of the saturation moment of $RuEu_{2-x}Ce_xSr_2Cu_2O_{10}$.

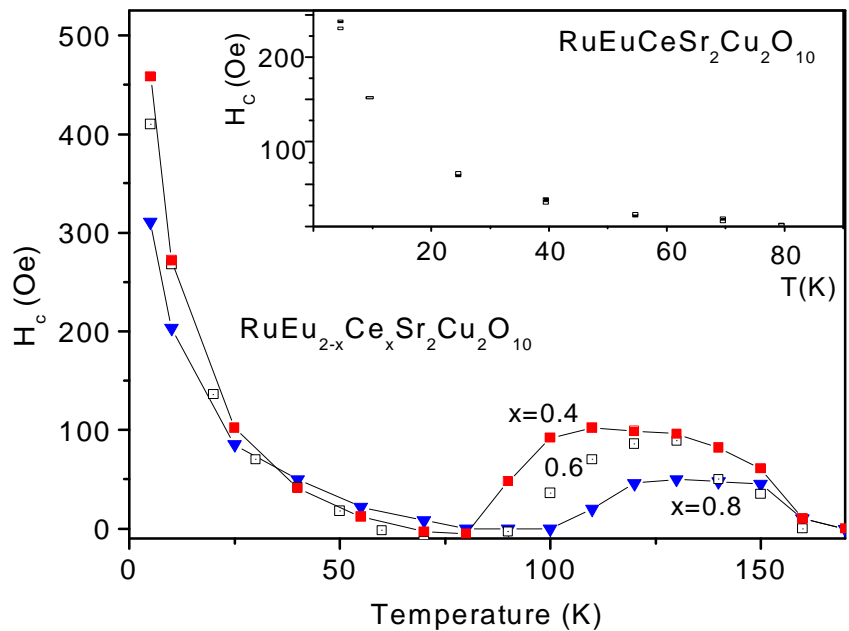

FIG. 6. The temperature dependence of the coercive fields ($H_C$) for $RuEu_{2-x}Ce_xSr_2Cu_2O_{10}$. $H_C$ curves for asp (bold) and quenched (open) $RuEuCeSr_2Cu_2O_{10}$ are shown in the inset

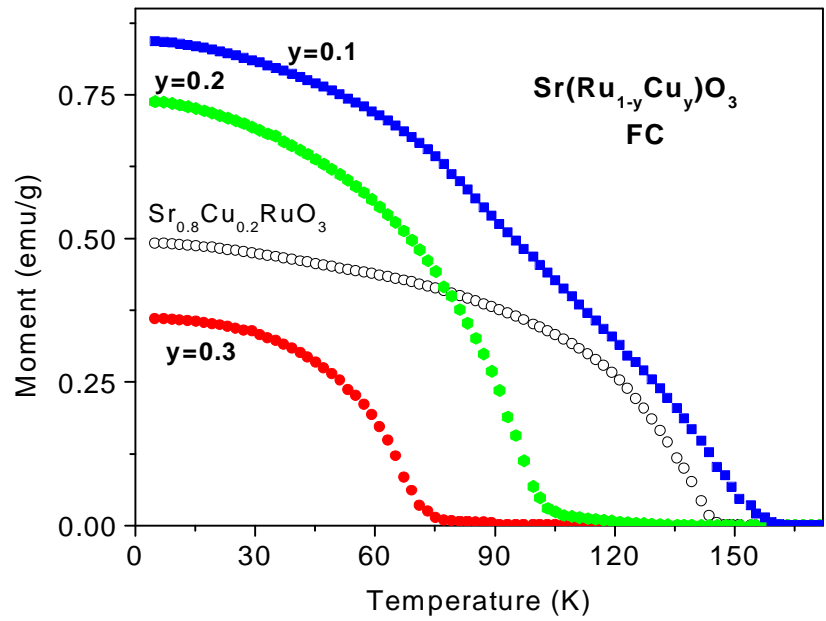

FIG. 7. FC magnetization curves (measured at 3 Oe) for $Sr_{0.8}Cu_{0.2}RuO_3$ (open symbols) and for $SrRu_{1-y}Cu_yO_3$ (y=0.1, 0.2 and 0.3).

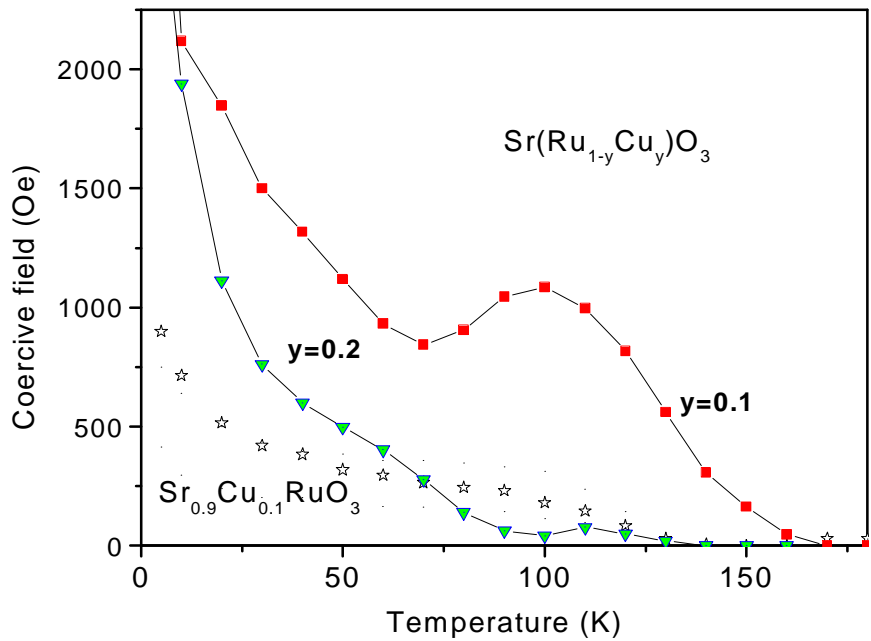

FIG. 8 The temperature dependence of the coercive fields for Sr-Cu-Ru-$O_3$ systems

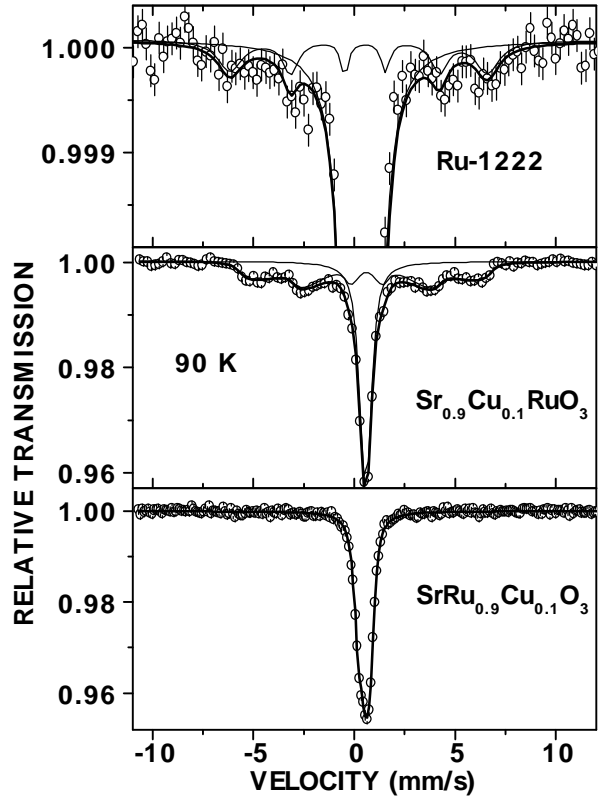

FIG. 9. Mössbauer spectra of SrRu$_{0.9}$Cu$_{0.1}$O$_3$, Sr$_{0.9}$Cu$_{0.1}$RuO$_3$ and Ru-1222 at 90 K.

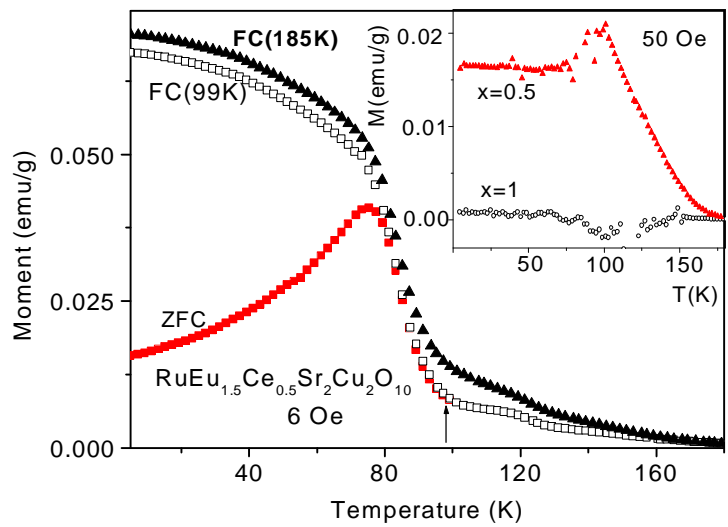

FIG.10 ZFC and FC (see text) magnetization curves of RuEu$_{1.5}$Ce$_{0.5}$Sr$_2$Cu$_2$O$_{10}$. The difference between the two FC curves (at 50 Oe) for x=1 and x=0.5 is shown in the inset.